\documentclass[conference]{IEEEtran}

\usepackage[utf8]{inputenc} 
\usepackage[T1]{fontenc}
\usepackage{url}
\usepackage{cite}
\usepackage[cmex10]{amsmath} 
\usepackage{amsfonts}
\usepackage{amssymb}
\usepackage{mathtools}
\usepackage{graphicx}
\usepackage{bm}

\usepackage{xfrac}
\usepackage{epstopdf}
\usepackage[font=small,labelfont=bf,justification=centering]{caption}
\usepackage[caption=false,font=footnotesize]{subfig}
\def\BibTeX{{\rm B\kern-.05em{\sc i\kern-.025em b}\kern-.08em
    T\kern-.1667em\lower.7ex\hbox{E}\kern-.125emX}}
\graphicspath{{.}{../eps/}{./images/}{./images/result/}}
\begin{document}

\title{Entropy Rate of Time-Varying Wireless Networks}

\author{\IEEEauthorblockN{Arta Cika\IEEEauthorrefmark{1}, Mihai-Alin Badiu\IEEEauthorrefmark{1}\IEEEauthorrefmark{2}, Justin P. Coon\IEEEauthorrefmark{1}, Shahriar Etemadi Tajbakhsh\IEEEauthorrefmark{1}}
\IEEEauthorblockA{\IEEEauthorrefmark{1}
Department of Engineering Science\\
University of Oxford, Parks Road, Oxford, UK, OX1 3PJ\\
Email: \{arta.cika and justin.coon\}@eng.ox.ac.uk}
\IEEEauthorblockA{\IEEEauthorrefmark{2}
Department of Electronic Systems\\
Aalborg University, Fredrik Bajers Vej 7, 9220 Aalborg {\O}st, Denmark}}

\maketitle

\begin{abstract}
In this paper, we present a detailed framework to analyze the evolution of the random topology of a time-varying wireless network via the information theoretic notion of entropy rate. We consider a propagation channel varying over time with random node positions in a closed space and Rayleigh fading affecting the connections between nodes. The existence of an edge between two nodes at given locations is modeled by a Markov chain, enabling memory effects in network dynamics. We then derive a lower and an upper bound on the entropy rate of the spatiotemporal network. The entropy rate measures the shortest per-step description of the stationary stochastic process defining the state of the wireless system and depends both on the maximum Doppler shift and the path loss exponent. It characterizes the topological uncertainty of the wireless network and quantifies how quickly the underlying topology is varying with time.
\end{abstract}

\begin{IEEEkeywords} 
Entropy rate, graph entropy, network topology, random geometric graphs.
\end{IEEEkeywords}

\section{Introduction}

Many complex systems show spatiotemporal characteristics in the real world. These networks are composed of a large number of nodes embedded in space and are dynamically evolving over time~\cite{newman2003structure, boccaletti2006complex}. Nodes in these networks are not isolated single elements, but they communicate with each other. These interactions can be represented as edges in a graph. In these systems, a node's state parameters (its healthy or infected condition, its availability or congestion) depend on both time and space-related variables~\cite{williams2016spatio}. Spatial networks~\cite{barthelemy2011spatial} that show time-dependence can be described as a time-ordered sequence of static graphs~\cite{holme2015modern}. Each graph is composed of a set of vertices (or nodes) and a set of edges linking nodes together. The positions of the nodes and edges of the graph change over time, therefore the topological structure shows a time-dependent evolution. Understanding the evolution of the links and quantifying the topological uncertainty is beneficial for modeling dissemination of information~\cite{cheng2013diffusion}, network synchronization~\cite{han2014robust}, routing protocols~\cite{pelusi2006opportunistic}, evaluation of route stability~\cite{an2002entropy} and fault localization~\cite{lgorzata2004survey}.

Entropy-based methods have been widely used to better understand the properties of complex networks. Dehmer et al.~\cite{dehmer2008information} introduced an information functional on each of the vertices of a graph in order to associate an entropic quantity with the network. The authors in~\cite{tee2017vertex} used local vertex measures of entropy to identify critical nodes in a network. In \cite{anand2009entropy,anand2011shannon} Anand et al. studied the Shannon and von Neumann entropy of graph ensembles. In the field of temporal networks, entropy rates of random walks have been used to characterize the temporal network structure~\cite{rosvall2014memory, saramaki2015exploring}. More recently, studies in~\cite{coon2016topological, coon2017topological, cika2017effects, badiu2018distribution} used Shannon entropy to quantify the topological uncertainty of wireless networks embedded within a spatial domain and the authors in~\cite{coon2018conditional} derived different lower bounds on the Shannon entropy of random geometric graphs by using the notion of conditional entropy. 

The typical approach used in the literature to characterize the entropy of the spatial networks has been limited to the study of the system as a static network. This approach has a fundamental limitation in the analysis of the evolution of the interactions, as it neglects any temporal property of the network. For instance, in wireless networks, links change over time due to the ability of the nodes to move inside the confined space and/or due to the variations in the propagation channel. In this environment, connections between nodes are established and broken intermittently, leading to frequently-changing network topologies. 

From an information theoretic perspective, the entropy rate measures the average minimum description length of a stationary stochastic process capturing the state of the dynamic system~\cite{cover2012elements}. Therefore a high entropy rate indicates a large uncertainty, or easiness of communication between nodes~\cite{gomez2008entropy}. As time goes on, two nodes initially disconnected might eventually establish a link between them, and hence a successful communication can take place. On the other hand, a high entropy rate indicates that the topology is frequently changing over time and nodes need to constantly send updated information about their state to the neighbors, leading to an increase in overhead throughout the network.

In this paper, we model a wireless network as a soft random geometric graph~\cite{penrose2016connectivity} evolving over time. Our model considers random node positions bounded inside a square/circle/triangle, small-scale fading affecting connections between nodes, and moving scatterers. The transmission range of each node and the quality of the channel jointly affect link connectivity. The availability of a link between two nodes is modeled as a two-state Markov chain with on and off states\cite{wang1995finite, zhang1999finite}. The main contributions of this paper are the modeling of the network state as a stationary stochastic process, the analyzing of the evolution of the topology via entropy rate and the derivation of lower and upper bounds on the entropy rate of the spatiotemporal network. To the best of our knowledge, this is the first time such analysis has been performed. 

This paper is the first research work to be included in a bigger project whose goal is to understand dynamics and mobility in spatiotemporal networks. Future work will focus on the analysis of the system stability via the information theory notion of entropy rate. Finally, we will use this metric for the optimization of routing protocols in wireless networks.
 
The remainder of the paper is organized as follows. In Section~\ref{sec:system_model} we explain the details of the network under examination and introduce the two-state Markov Model used to characterize the evolution of links. We provide a lower and an upper bound on the shortest description of the stochastic process by analyzing the entropy rate of the dynamic system in Section~\ref{sec:entropy_rate}. We then present numerical results on the tightness of the derived entropy bounds for nodes embedded in a square/circle/triangle in Section~\ref{sec:numerical}. Finally, in Section~\ref{sec:conclusions}, we present our conclusions.
\section{System model}\label{sec:system_model}
\subsection{Network Model}\label{sec:network_model}
We consider a network of $\mathcal{V}_n=\{1,\dotsc,n\}$ of $n$ nodes representing wireless devices located randomly in a space $\mathcal{K}\subset\mathbb{R}^d$ where $d\in \mathbb{N}$ with $d\ge2$, with finite volume and diameter $D\coloneqq\text{sup}_{u,v\in\mathcal{K}}\| u-v\|$. The locations of the nodes, $\mathbf{Z}\coloneqq(Z_{i})_{i\in\mathcal{V}_n}$ are independently and uniformly distributed in $\mathcal{K}$. We denote by $\mathbf{R}\coloneqq(R_{i,j})_{i<j}$ the random vector collecting the pair distances $R_{i,j} = \|Z_i-Z_j\|$, and let $f_{\mathbf{R}}: [0,D]^{n(n-1)/2}\rightarrow[0,\infty)$ be its pdf.

We model a wireless network as an instance of a soft undirected random geometric graph (RGG). We represent the time-varying wireless network as a time-ordered sequence of RGGs. Each graph corresponds to a snapshot of the wireless network at a particular time instance. Let $X_{i,j}^t$ be a Bernoulli random variable that models the existence of the edge $\left(i,j\right)$ between nodes $i$ and $j$ at time $t$, and define $\mathbf{X}^t=\left(X_{i,j}^t\right)_{i<j}$ to be the state of the network at time $t$. In the next subsection, we model the existence of an edge between two nodes at given locations as a two-state Markov chain with on and off states. This enables us to derive a lower and upper bound on the entropy rate of the dynamic system, as shown in section~\ref{sec:entropy_rate}.
\subsection{Two-State Markov Model for Connection Links}\label{sec:markov_model}
In a real-world wireless communication system, the environment can cause the creation of several reflected, diffracted, and/or scattered copies of a transmitted signal. Therefore, the receiving antenna picks up a superposition of a series of attenuated and delayed versions of the original signal. A transmission from node $i$ to node $j$ is successful if the received instantaneous signal-to-noise ratio $\left(\text{SNR}\right)$, $\gamma_{i,j}$, is greater than a certain threshold $\gamma_{th}$ determined by the communication hardware, and the modulation and coding scheme of the wireless system. Therefore, at every time-step $t$, the Bernoulli random variable $X^t_{i,j}$, conditioned on the pair distance $R_{i,j}$, is equal to
\begin{equation}\label{eq:bernoulli_rv}
X_{i,j}^t | R_{i,j}=
\begin{cases}
0, & \text{if } 0<\gamma^t_{i,j} < \gamma_{th}, \\
1, & \text{if } \gamma_{th} \le \gamma^t_{i,j} < \infty.
\end{cases}
\end{equation}
In a typical multipath propagation environment, the received signal envelope shows a Rayleigh distribution. With additive Gaussian noise, the received instantaneous SNR $\gamma_{i,j}$ has an exponential distribution with probability density function
\begin{equation}\label{eq:pdf_snr}
f(\gamma_{i,j})=\frac{1}{\gamma_0}\mathrm{e}^{-\sfrac{\gamma_{i,j}}{\gamma_0}}, \quad \gamma_{i,j}\geq 0
\end{equation}
where $\gamma_0$ is the average SNR, i.e. $\gamma_0=E\left[\gamma_{i,j}\right]$. The SNR at the receiver (in the absence of interference) decays with distance like $\gamma_0 \sim r_{i,j}^{-\eta}$, where $\eta>0$ is the path loss exponent. Hence, the probability of existence of the edge $\left(i,j\right)$, conditioned on the pair distance, at time $t$, is equal to   
\begin{align}\label{eq:P_edge_existence}
\mathbb{P}\left(X_{i,j}^t=1|R_{i,j}=r_{i,j}\right)&=\mathrm{e}^{-\sfrac{\gamma_{th}}{\gamma_0}}\nonumber\\ 
&=\mathrm{e}^{-\left(\sfrac{r_{i,j}}{r_0}\right)^\eta}
\end{align}
where $r_0\sim\left(\sfrac{1}{\gamma_{th}}\right)^{\sfrac{1}{\eta}}$ defines the typical connection range and depends on several system parameters, such as the transmit power, wavelength, bandwidth and the noise spectral density. Typically, eq.~\eqref{eq:P_edge_existence} is referred to as the pair connection function for nodes $i$ and $j$. 

In a time-varying multipath channel, the emitted wave is subject to the Doppler effect, i.e., a shift in the received frequency due to the movements of either the transmitter, the receiver, and/or external scatterers. In this work, we assume stationary terminals and moving scatterers~\cite{andersen2009doppler}. Extensions to mobile terminals are left to future work. The fading characteristics of the channel are determined by the maximum Doppler frequency, $\nu_{i,j}$, which is a measure for the rate of change of the channel\cite{molisch2012wireless}. The level crossing rate of the instantaneous received SNR $\gamma_{i,j}$ is a measure of the rapidity of the fading. It quantifies how often the signal level crosses the threshold $\gamma_{th}$, usually in the positive-going direction, and is defined as~\cite{wang1995finite} 
\begin{equation}
\text{LCR}\left(\gamma_{th}\right)=\sqrt{2\pi}\left(\frac{\gamma_{th}}{\gamma_0}\right)^{\sfrac{1}{2}}\ \nu_{i,j} \ \mathrm{e}^{-\sfrac{\gamma_{th}}{\gamma_0}}.
\end{equation}
Expressing the level crossing rate as a function of pair distance $R_{i,j}$, we obtain
\begin{equation}
\text{LCR}\left(r_{i,j}\right)=\sqrt{2\pi}\left(\frac{r_{i,j}}{r_0}\right)^{\sfrac{\eta}{2}}\ \nu_{i,j} \ \mathrm{e}^{-\left(\sfrac{r_{i,j}}{r_0}\right)^\eta}.
\end{equation}
Under the slow fading assumption, the level crossing rate at $\gamma_{th}$ is much smaller than the average number of symbols per second transmitted when the channel is in state ``on'' (a link exists) or ``off'' (a link does not exist). Hence, in a communication system with a transmission rate of $B$ symbols per second,  the state transition probability, conditioned on the pair distance $R_{i,j}$, can be approximated as~\cite{wang1995finite}
\begin{multline}\label{eq:Pa,1-a}
\mathbb{P}\left(X_{i,j}^{t}=1-a|X_{i,j}^{t-1}=a, R_{i,j}=r_{i,j}\right)\\
\approx\frac{\text{LCR}(r_{i,j})}{\mathbb{P}\left(X_{i,j}^{t-1}=a|R_{i,j}\right)\times B}
\end{multline}
for each $a\in\{0,1\}$. Conditioned on the pair distances, we assume that the edge trajectories, $X_{i,j}^1,\dotsc,X_{i,j}^t$, $\forall i<j$, are independent and we model the evolution of each edge $\left(i,j\right)$ as a stationary Markov chain. That is,
\begin{multline}\label{eq:network_state_conditioned}
\mathbb{P}\left(\mathbf{X}^1=\mathbf{x}^1,\dotsc,\mathbf{X}^t=\mathbf{x}^t|\mathbf{R}=\mathbf{r}\right)\\
=\prod_{i<j}\left[\prod_{u=2}^t\mathbb{P}\left(X_{i,j}^u=x_{i,j}^{u}|X_{i,j}^{u-1}=x_{i,j}^{u-1},R_{i,j}=r_{i,j}\right)\right.\\
\left.\times\mathbb{P}\left(X_{i,j}^1=x^1_{i,j}|R_{i,j}=r_{i,j}\right)\vphantom{\int}\right]
\end{multline}
for each $\mathbf{x}^u\in\{0,1\}^{n(n-1)/2}, u=1,\dotsc,t$ and $\mathbf{r}\in[0,D]^{n(n-1)/2}$. Therefore, the joint pdf of the network state variables is obtained by averaging eq.~\eqref{eq:network_state_conditioned} over the pair distances, i.e.,
\begin{align}
&\mathbb{P}\left(\mathbf{X}^1=\mathbf{x}^1,\dotsc,\mathbf{X}^t=\mathbf{x}^t\right)\nonumber\\
&=\int_{\mathcal{R}}\mathbb{P}\left(\mathbf{X}^1=\mathbf{x}^1,\dotsc,\mathbf{X}^t=\mathbf{x}^t|\mathbf{R}=\mathbf{r}\right)f_{\mathbf{R}}(\mathbf{r})\text{d}{\mathbf{r}}
\end{align}
where the integration domain is $\mathcal{R}=[0,D]^{n(n-1)/2}$. After averaging, the resulting stochastic process $\mathbf{X}^t$ does not possess the Markov property. However, $\mathbf{X}^t$ inherit the stationary property, i.e., the joint distribution of any subset of the sequence of the random variables is invariant with respect to shifts in the time index~\cite{cover2012elements}. In the next Section, we make progress by deriving a lower and an upper bound on the entropy rate of the dynamic system. 
\section{Entropy Rate of a Spatiotemporal Network} \label{sec:entropy_rate}
We quantify the topological uncertainty in the context of wireless networks by studying the entropy rate of the stochastic process capturing the state of the dynamic system. The entropy rate of the stationary stochastic process $\mathbf{X}^t=(X_{i,j}^t)_{i<j}$ is defined by~\cite{cover2012elements}
\begin{equation}
H\left(\mathcal{X}\right)=\lim_{t\to\infty}\frac{1}{t}H\left(\mathbf{ X}^{1},\dotsc, \mathbf{ X}^{t}\right).
\end{equation}
We can interpret the entropy rate as a measure of the uncertainty of the future state of the dynamic system given its past states. It quantifies how quickly the underlying network topology is varying with time.
\subsection{Upper Bound on the Entropy Rate}
In the following, we consider the trajectories of individual links, i.e., $X_{i,j}^1,\dotsc,X_{i,j}^t$, $\forall i<j$. For any time-step $t$, we can write
\begin{equation}\label{eq:subadditivity}
H\left(\mathbf{ X}^{1},\dotsc, \mathbf{ X}^{t}\right)\le \sum_{i<j}H\left(X_{i,j}^{1},\dotsc, X_{i,j}^{t}\right)
\end{equation}
where the inequality follows from the subadditivity property of the entropy, i.e., the joint entropy can not be greater than the sum of the entropies of disjoint subsets of variables. Equality holds if and only if $\left(X_{i,j}^t\right)_{i<j}$ are independent. Next, by the chain rule, it follows that
\begin{align}
&H\left(\mathbf{ X}^{1},\dotsc, \mathbf{ X}^{t}\right)\nonumber \\
&\le \sum_{i<j}\left[\sum_{u=2}^tH\left(X_{i,j}^{u}| X_{i,j}^{u-1},\dotsc, X_{i,j}^{1}\right) + H\left(X_{i,j}^{1}\right)\right].
\end{align}
A fundamental result of information theory states that conditioning reduces entropy~\cite{cover2012elements}. Hence, we can write
\begin{align}\label{eq:stationarity}
H\left(\mathbf{ X}^{1},\dotsc, \mathbf{ X}^{t}\right)&\leq \sum_{i<j}\left[\sum_{u=2}^t H\left(X_{i,j}^{u}| X_{i,j}^{u-1}\right) + H\left(X_{i,j}^{1}\right)\right]\nonumber\\ 
&=\sum_{i<j}\left[(t-1)H\left(X_{i,j}^{2}| X_{i,j}^{1}\right) + H\left(X_{i,j}^{1}\right)\right]
\end{align}
where the second equality follows from the stationary property of the stochastic process $\mathbf{X}^t=(X_{i,j}^t)_{i<j}$. Dividing by $t$ and taking the limit $t\rightarrow\infty$, and by using eqs.~\eqref{eq:subadditivity} and~\eqref{eq:stationarity}, we arrive at the entropy rate relation
\begin{equation}\label{eq:upper_bound}
H\left(\mathcal{X}\right)\leq\sum_{i<j}H\left(X_{i,j}^{2}| X_{i,j}^{1}\right)
\end{equation}
where
\begin{align}\label{eq:cond_entropy}
&H\left(X_{i,j}^2| X_{i,j}^1\right) = -\sum_{a\in\{0,1\}}\mathbb{P}\left(X_{i,j}^1=a\right)\nonumber\\
&\times  \sum_{b\in\{0,1\}} \mathbb{P}\left(X_{i,j}^2=b |X_{i,j}^1=a\right)\log \mathbb{P}\left(X_{i,j}^2=b |X_{i,j}^1=a\right).
\end{align}
$\mathbb{P}\left(X_{i,j}=a\right)$ is simply the probability that edge $(i,j)$ exists $\left(a=1\right)$ or not $\left(a=0\right)$, averaged over the pair distance $R_{i,j}$. More accurately, we can write
\begin{align}\label{eq:average_edge_prob}
&\mathbb{P}\left(X_{i,j}^1=a\right)\nonumber\\
&=\int_0^D\mathbb{P}\left(X_{i,j}^1=a | R_{i,j}=r_{i,j}\right)f_{R_{i,j}}\left(r_{i,j}\right)\text{d}r_{i,j}.
\end{align}
It is straightforward to notice that these probabilities are equal for different links, i.e., $P\left(X_{i,j} = a\right) = P\left(X_{k,l} = a\right)$ for $\left(k,l\right)\neq \left(i,j\right)$.  In the same fashion, $\mathbb{P}\left(X_{i,j}^2=b |X_{i,j}^1=a\right)$ is the edge transition probability averaged over the pair distance, and is given by
\begin{align}\label{eq:average_edge_trans_prob}
&\mathbb{P}\left(X_{i,j}^2=b | X_{i,j}^1=a\right)\nonumber\\
&=\int_{0}^D\mathbb{P}\left(X_{i,j}^2=b | X_{i,j}^1=a, R_{i,j}=r_{i,j}\right)f_{R_{i,j}}\left(r_{i,j}\right)\text{d}r_{i,j}.
\end{align}
It is clear from eqs.~\eqref{eq:Pa,1-a} and~\eqref{eq:average_edge_trans_prob} that when different links have equal maximum Doppler frequencies, they will also have equal transition probabilities. Assuming that the maximum Doppler frequencies of different links are equal, the upper bound on the entropy rate simplifies to
\begin{equation}\label{eq:upper_bound_entropy_rate}
H\left(\mathcal{X}\right)\leq\binom{n}{2}H\left(X_{1,2}^{2}| X_{1,2}^{1}\right).
\end{equation}
In the large $n$ limit the upper bound on the entropy rate  scales like $\mathcal{O}\left(n^2\right)$.
\subsection{Lower Bound on the Entropy Rate}
To enable us to find a lower bound on the entropy rate of a soft RGG evolving over time, we turn to the information theoretic notion of conditional entropy. By using a similar argument to that employed in eq.~\eqref{eq:stationarity}, it is possible to show that
\begin{equation}
H\left(\mathbf{ X}^{1},\mathbf{ X}^{2},\dotsc, \mathbf{ X}^{t}\right)\ge H\left(\mathbf{ X}^{1},\mathbf{ X}^{2},\dotsc, \mathbf{ X}^{t}|\mathbf{R}\right).
\end{equation}
Conditioned on pair distances, the link trajectories are independent, and each of them is a stationary Markov chain, as explained in section~\ref{sec:markov_model}. This leads naturally to the following entropy relation
\begin{align}
&H\left(\mathbf{ X}^{1},\mathbf{ X}^{2},\dotsc, \mathbf{ X}^{t}\right)\nonumber\\
&\geq\sum_{i<j}\left[(t-1)H\left(X_{i,j}^{2}| X_{i,j}^{1},R_{i,j}\right) + H\left(X_{i,j}^{1}|R_{i,j}\right)\right].
\end{align}
Equivalently, combining the arguments employed in eq.~\eqref{eq:upper_bound} with the assumption made in eq.~\eqref{eq:upper_bound_entropy_rate}, we obtain
\begin{equation}\label{eq:lower_boundeq_entropy_rate}
H(\mathcal{X})\ge\binom{n}{2}H\left(X_{1,2}^{2}| X_{1,2}^{1},R_{1,2}\right)
\end{equation}
where
\begin{align}\label{conditional_entropy_rate}
&H\left(X_{1,2}^{2}| X_{1,2}^{1},R_{1,2}\right) = -\int_{0}^D f_{R}(r_{1,2})\text{d}r_{1,2}\nonumber \\
&\times\sum_{a\in \{0,1\}}\mathbb{P}\left(X_{1,2}^1=a | R_{1,2}=r_{1,2}\right)\nonumber \\
&\times\sum_{b\in \{0,1\}}\left[\mathbb{P}\left(X_{1,2}^2=b | X_{1,2}^1=a, R_{1,2}=r_{1,2}\right)\right. \nonumber \\
&\left.\times\log\mathbb{P}\left(X_{1,2}^2=b | X_{1,2}^1=a, R_{1,2}=r_{1,2}\right)\right].
\end{align}
Hence, to evaluate the lower bound, we assume prior knowledge of the node locations and average the entropy rate over the spatial distribution. Putting together eqs.~\eqref{eq:upper_bound_entropy_rate},~\eqref{eq:lower_boundeq_entropy_rate}, and~\eqref{conditional_entropy_rate} manifests in the relation
\begin{equation}
\binom{n}{2}H\left(X_{1,2}^{2}| X_{1,2}^{1},R_{1,2}\right)\leq H(\mathcal{X})\leq \binom{n}{2}H\left(X_{1,2}^{2}| X_{1,2}^{1}\right).
\end{equation}

\begin{figure*}[!t]
\centering
\subfloat[]{\includegraphics[width=2.33in]{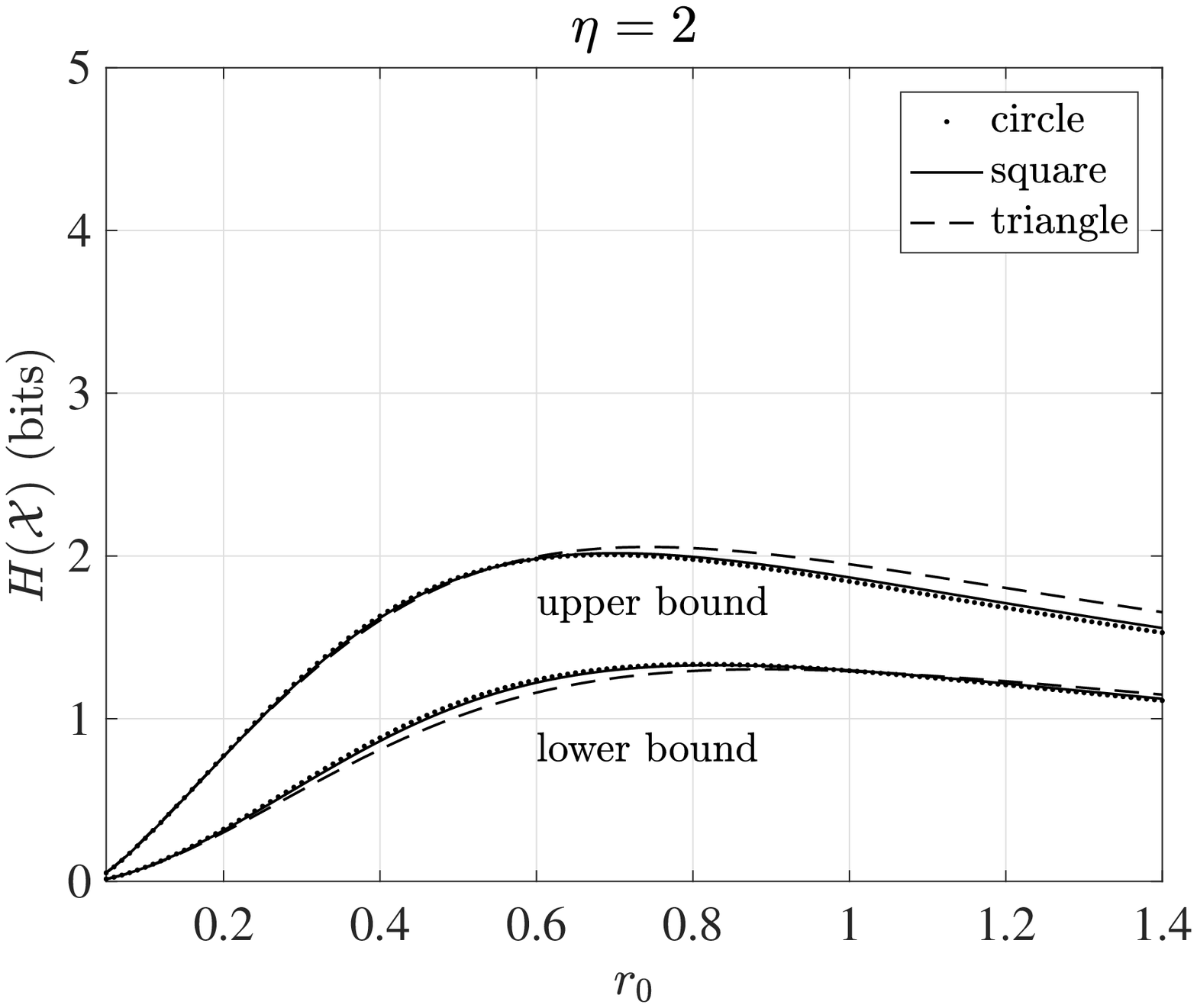}
\label{fig_entropy_rate_vs_r0_eta2}}
\hfil
\subfloat[]{\includegraphics[width=2.33in]{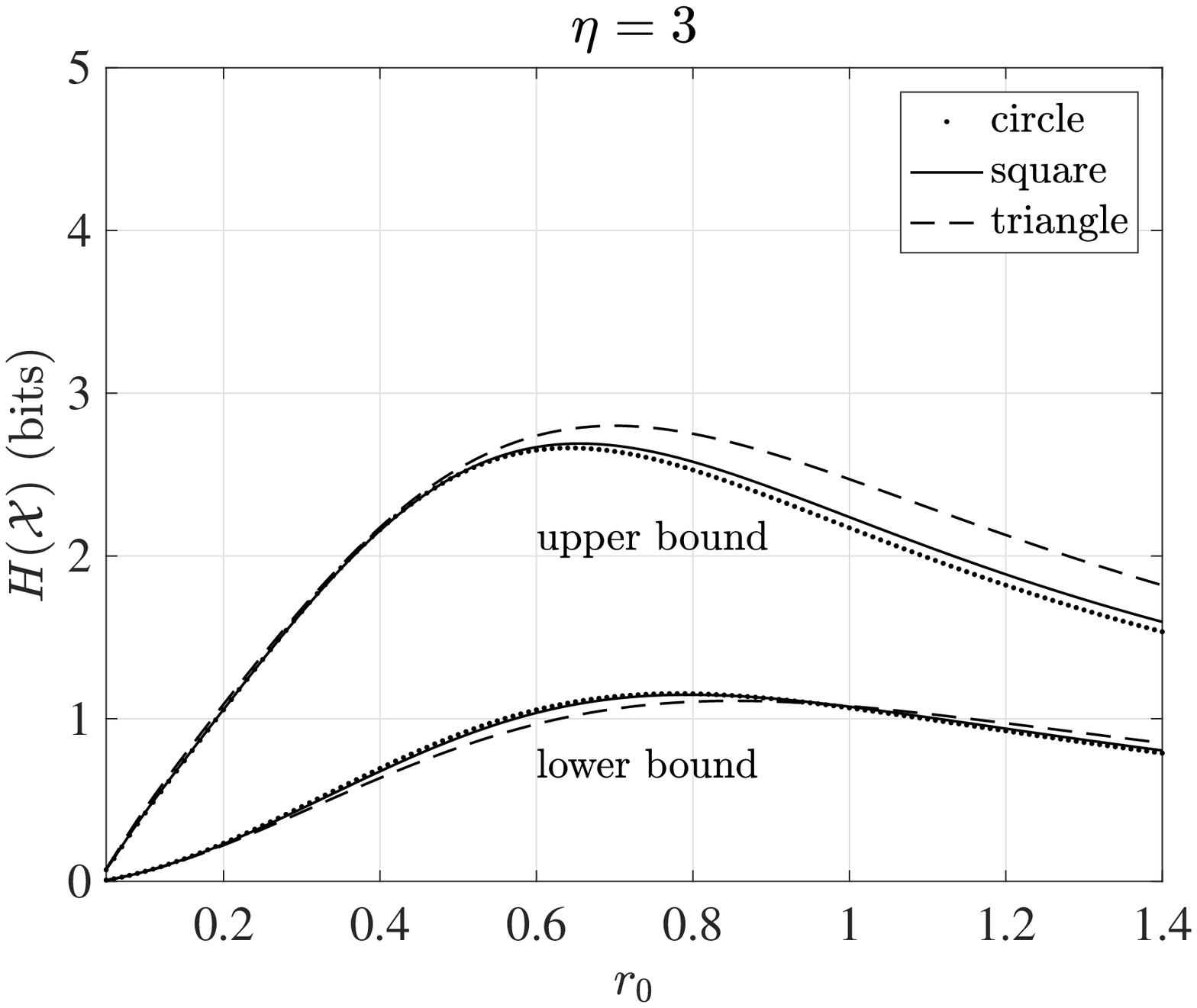}
\label{fig_entropy_rate_vs_r0_eta3}}
\hfil
\subfloat[]{\includegraphics[width=2.33in]{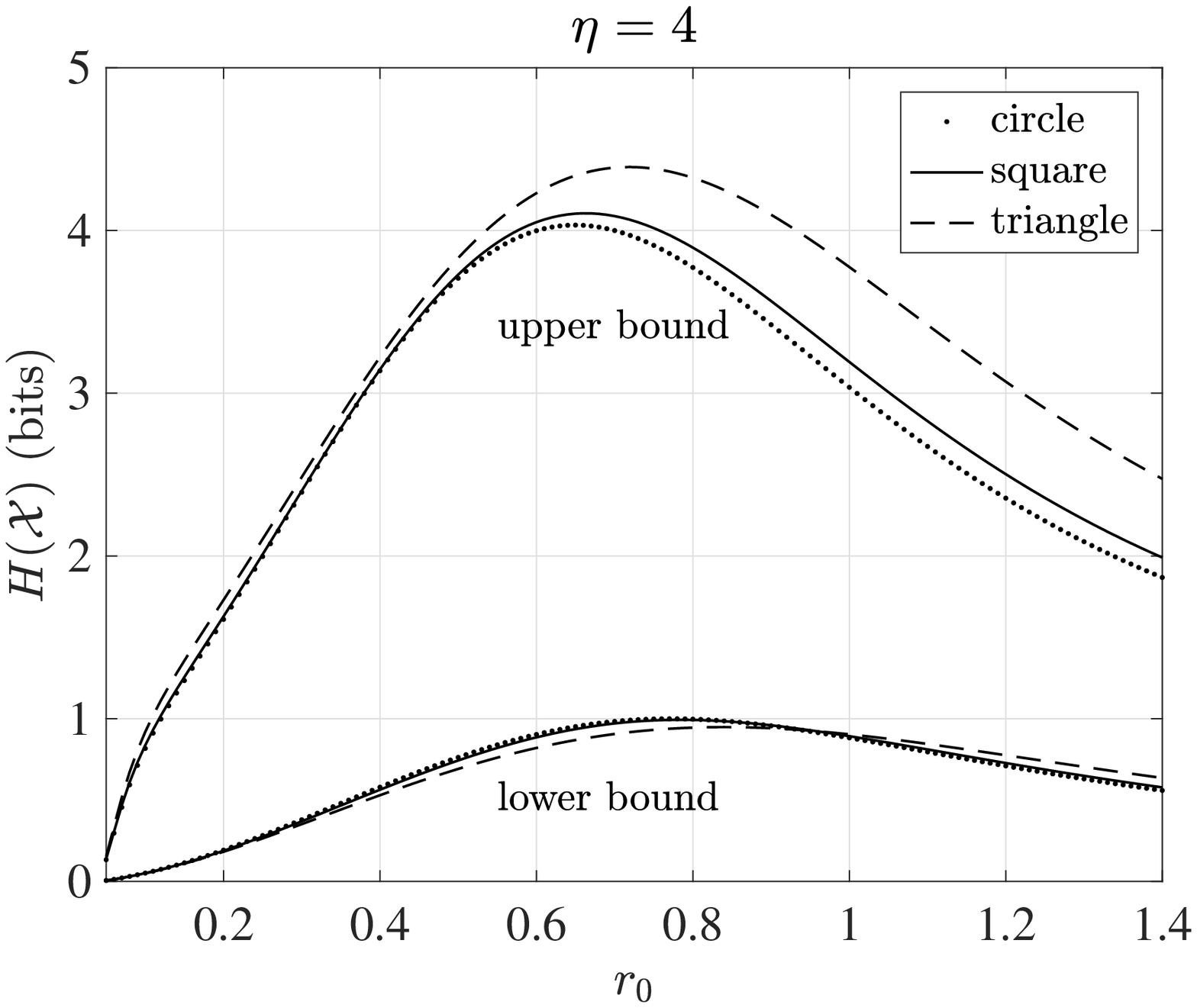}
\label{fig_entropy_rate_vs_r0_eta4}}
\caption{Entropy rate of a fifty-node RGG with soft pair connection function in two dimensions; bounding geometries:  square of unit side length, circle of radius $\sfrac{1}{\sqrt{\pi}}$, and equilateral triangle of side length $\sfrac{2}{\sqrt[4]{3}}$; path loss exponent values are $\eta = 2, 3, 4$; maximum Doppler frequency $\nu=500$ Hz. }
\label{fig:entropy_rate_vs_r0}
\end{figure*}
\begin{figure*}[!t]
\centering
\subfloat[]{\includegraphics[width=2.33in]{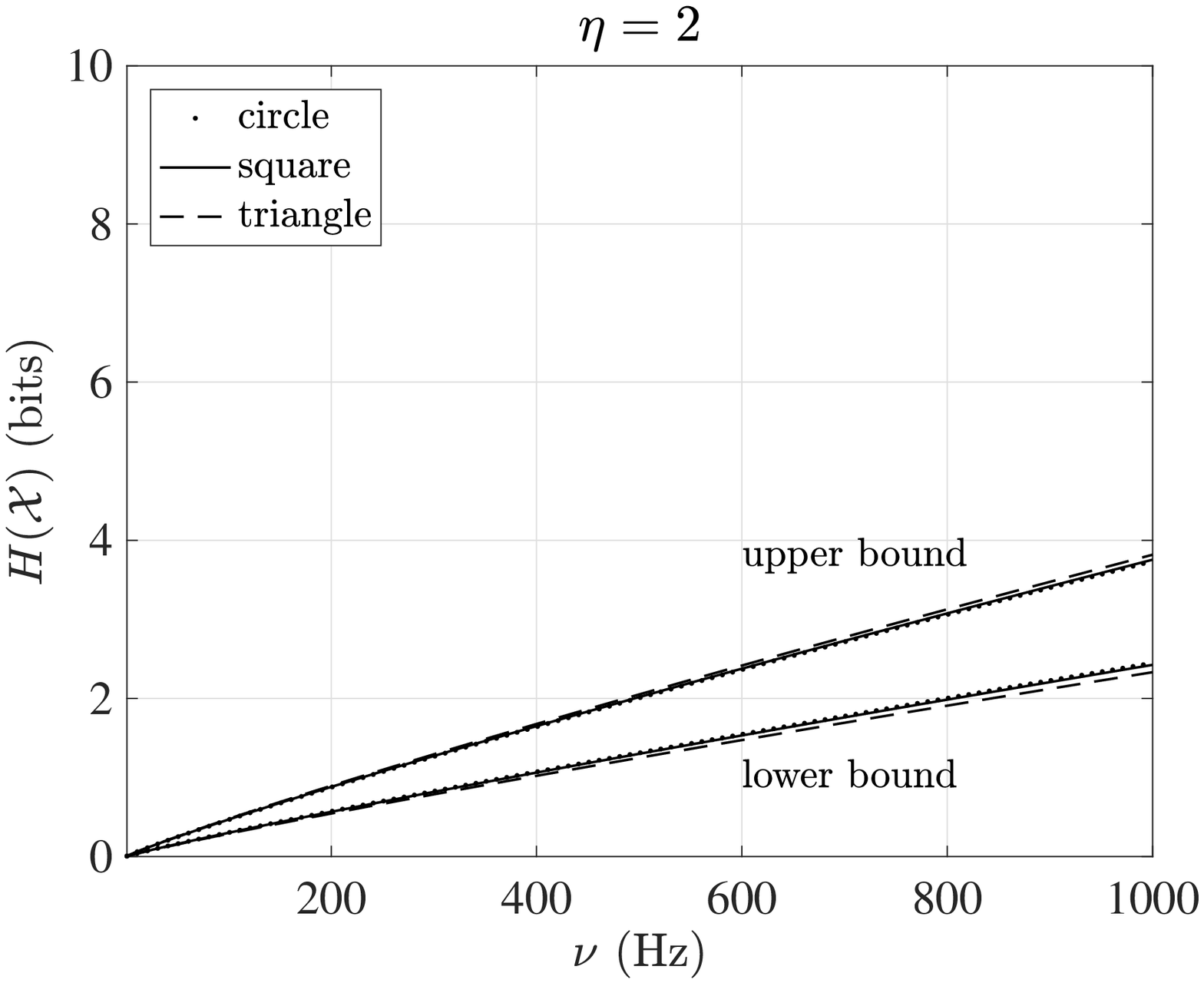}
\label{fig_entropy_rate_vs_f_eta2}}
\hfil
\subfloat[]{\includegraphics[width=2.33in]{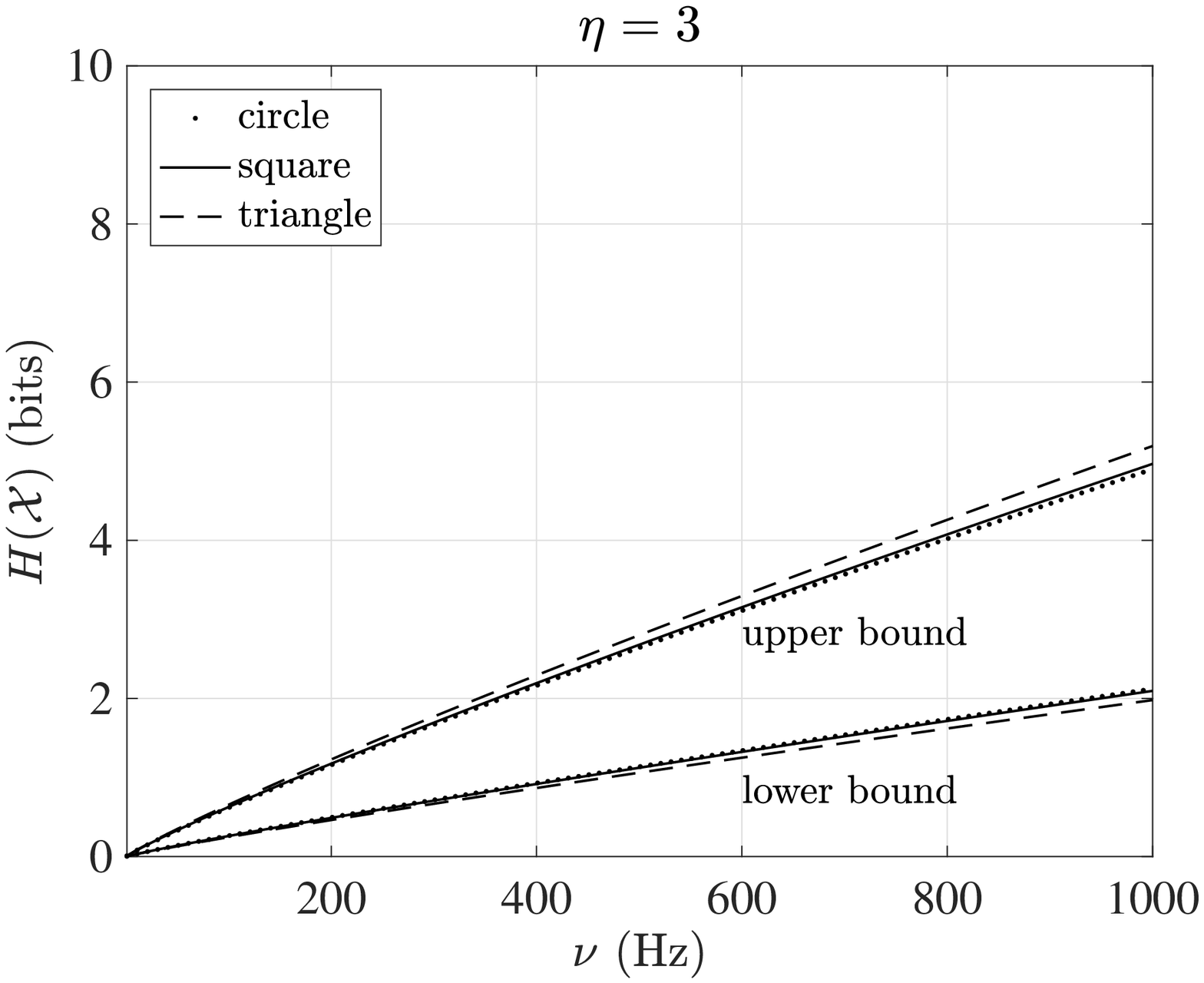}
\label{fig_entropy_rate_vs_f_eta3}}
\hfil
\subfloat[]{\includegraphics[width=2.33in]{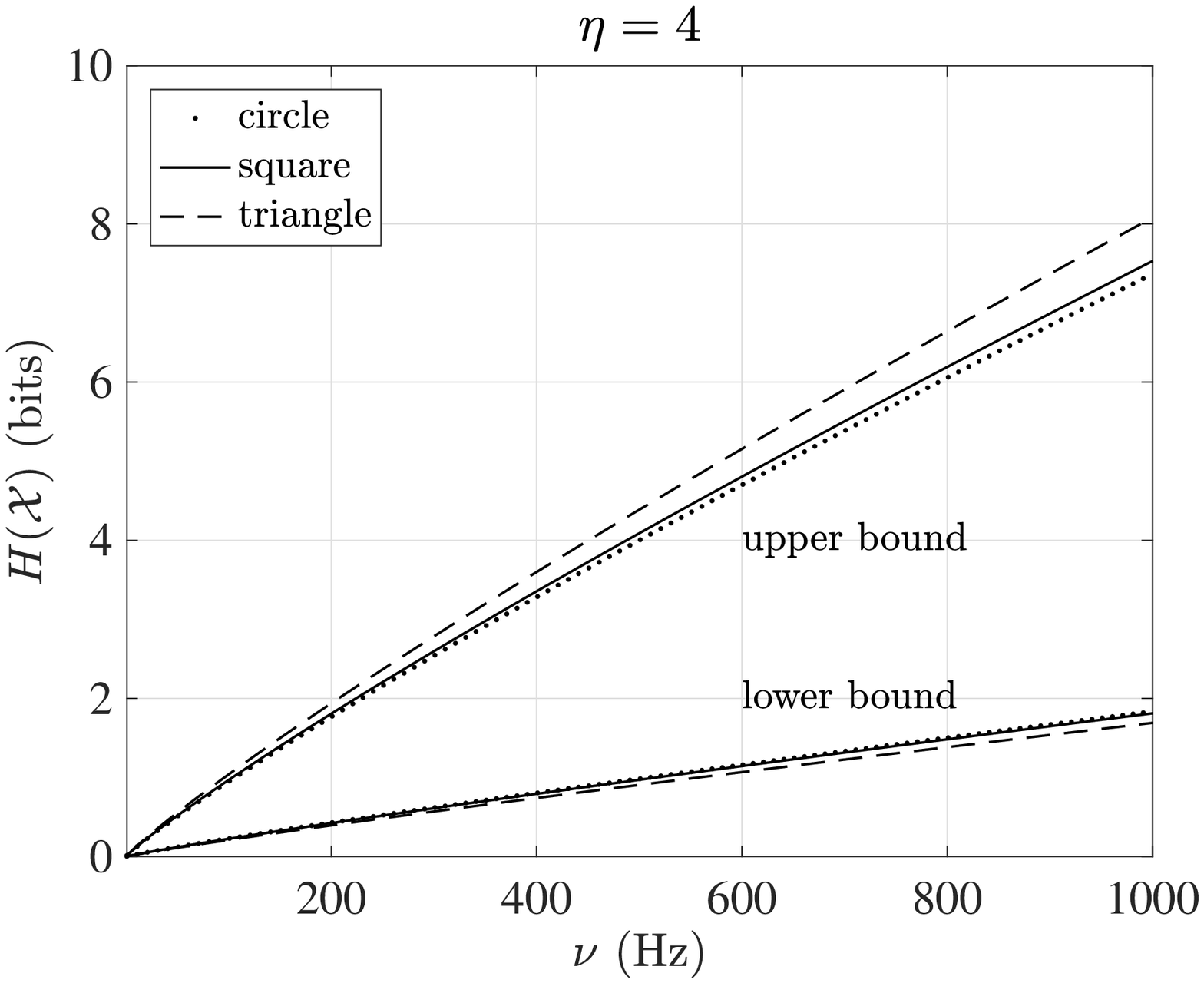}
\label{fig_entropy_rate_vs_f_eta4}}
\caption{Entropy rate of a fifty-node RGG with soft pair connection function in two dimensions; bounding geometries:  square of unit side length, circle of radius $\sfrac{1}{\sqrt{\pi}}$, and equilateral triangle of side length $\sfrac{2}{\sqrt[4]{3}}$; path loss exponent values are $\eta = 2, 3, 4$;  typical connection range $r_0=0.7$.}
\label{fig:entropy_rate_vs_f}
\end{figure*}
\section{Numerical Experiments}\label{sec:numerical}
In the following, we consider a network of $n=50$ nodes located randomly in a circle/square/triangle. The geometry of the domain $\mathcal{K}$ defines $f_{R}(r)$. Analytic expressions for $f_R(r)$ are known for simple, convex geometries~\cite{basel2014random,mathai1999introduction}. Before presenting the numerical results, we need to guarantee the assumptions of our model as stated in Section~\ref{sec:markov_model}. As aforesaid, we approximate the transition probabilities given in eq.~\eqref{eq:Pa,1-a} under the slow fading assumption. If we take, for instance, a communication system using 802.11a/g protocols, the symbol rate is $B=12$ MBd. Given this constraint, the maximum Doppler frequency $\nu$ must vary in the range $1\ \text{Hz}\leq \nu \leq 1\ \text{kHz}$ for $2\le \eta \le 5$.

We now study the entropy bounds derived in Section~\ref{sec:entropy_rate}. To that end, in Fig.~\ref{fig:entropy_rate_vs_r0}, we plot the upper and lower bound on the entropy rate of a fifty-node RGG for the cases where $\eta=2,3,4$ versus the typical connection range $r_0$. Then, we analyze the effect of the Doppler frequency on the entropy rate of the dynamic system. Fig.~\ref{fig:entropy_rate_vs_f} illustrates the entropy rate of a fifty-node RGG versus the maximum Doppler frequency, which is a measure for the rate of change of the channel, for different bounding geometries. If the maximum Doppler frequency increases, the topological uncertainty of network increases. A few important things can be noted from both Fig.~\ref{fig:entropy_rate_vs_r0} and Fig.~\ref{fig:entropy_rate_vs_f}. First, we notice that for practical values of the Doppler frequency the system's dynamics can be quantified by a few bits. Second, it can be observed that when the nodes are randomly located inside a square or a circle, the geometry of the domain does not significantly affect the entropy rate. Whereas, choosing triangle as the confining geometry has a non-negligible effect on it. Third, the increase in the upper bound and the reduction in lower bound for increasing ``hardness'' in the connection function~\eqref{eq:P_edge_existence} is evident in these figures. Therefore, we can notice that the gap between the upper and lower bound widen with increasing $\eta$. Mathematically, the parameter $\eta$ controls the stretch of the decaying exponential. For $\eta\rightarrow\infty$, we recover the hard connection model. In the case of the lower bound, as we increase $\eta$, the pairwise uncertainty (and consequently the conditional entropy rate) decreases~\cite{coon2017entropy}. Clearly, as $\eta\rightarrow\infty$ the lower bound tends to zero, but the upper bound on entropy rate $H(\mathcal{X})$ is still $\mathcal{O}(n^2)$. Finally, for very soft connection functions there is little dependence upon the spatial embedding. 
\section{Conclusions and Future Work}\label{sec:conclusions}
In this paper, we formally analyzed the evolution of the random topology of a time-varying wireless network via the entropy rate. In this regard, we took into account both the temporal and the spatial properties of the system. The entropy rate provides insights into the topological uncertainty of the spatiotemporal network and quantifies how quickly the underlying topology is varying with time. Given the pair distances, we modeled the evolution of each edge as a stationary Markov chain. We then showed that the stochastic process describing the state of the dynamic system is stationary. This result enabled us to derive a simple upper bound on the entropy rate. We also introduced the concept of conditional entropy rate and used it to derive a lower bound on the entropy rate. Both bounds are obtained under the assumption that the maximum Doppler frequencies of different links are equal. The results presented in this paper provide insights into how the time-varying fading channel affects the dynamics of the network. This perspective can be useful for storing and communicating network topology information, calculating routing tables and for fault localization and operational management in Software-defined networking. 
 
The presented framework can be extended to mobile ad-hoc wireless networks to analyze the topological uncertainty caused by node movement using different mobility models. Future work will focus on the analysis of the system stability through the concept of entropy rate, with the goal to develop a new metric to optimize routing protocols.
\section*{Acknowledgment}
The authors wish to acknowledge the support of Moogsoft and EPSRC under grant number EP/N002350/1 (``Spatially Embedded Networks''). Work by M.-A. Badiu was supported by Independent Research Fund Denmark grant number DFF-5054-002 and was carried out during his visit to University of Oxford.

\bibliographystyle{ieeetr}
\IEEEtriggeratref{24}
\bibliography{IEEEabrv,biblio}
\end{document}